\documentclass[12pt]{article}
\usepackage{graphicx}
\usepackage{amssymb}
\title{
Analog geometry in an expanding fluid from AdS/CFT perspective
 }
\author{
Neven Bili\'c, Silvije Domazet and Dijana Toli\'c\\
Division of Theoretical Physics, Rudjer Bo\v{s}kovi\'{c} Institute, \\
P.O.\ Box 180, 10001 Zagreb, Croatia \\
E-mail: bilic@irb.hr, silvije.domazet@irb.hr, dijana.tolic@irb.hr
}
\date{\today}
\begin{document}
\maketitle
\begin{abstract}

%
The dynamics of an expanding hadron fluid 
at temperatures below the chiral transition is studied in the framework of AdS/CFT correspondence.
We establish a correspondence between the asymptotic AdS geometry in the  the 4+1-dimensional bulk
with the analog spacetime geometry on its 3+1 dimensional  boundary 
with the background fluid undergoing a spherical
Bjorken type expansion.
The analog metric tensor on the boundary
depends locally on the
soft pion dispersion relation
and the four-velocity
 of the fluid. 
 The AdS/CFT correspondence provides a relation between the pion velocity
 and the critical temperature of the chiral phase transition.

\end{abstract}

%
%


\section{Introduction}

\label{introduction}

The original formulation of gauge-gravity duality in the form of the so called AdS/CFT correspondence establishes 
 an equivalence of a four dimensional ${\cal{N}}=4$ supersymmetric Yang-Mills theory
and string theory in a ten dimensional ${\rm AdS}_5\times {\rm S}_5$
bulk \cite{maldacena,gubser,witten1}.
However, the AdS/CFT correspondence goes beyond pure string theory as it links 
many other important theoretical and phenomenological issues such as
fluid dynamics \cite{rangamani}, 
  thermal field theories, black hole physics, quark-gluon plasma \cite{kovtun}, 
gravity and cosmology.
 In particular, the AdS/CFT correspondence proved to be  useful in
 studying some properties of 
strongly interacting matter  \cite{erlich}
described at the fundamental level by
a  theory called
 quantum chromodynamics (QCD), although ${\cal{N}}=4$ supersymmetric Yang-Mills theory
  differs substantially from QCD.
 
  Our purpose is to study in terms of the  AdS/CFT correspondence a class of field theories with spontaneously
  broken symmetry 
 restored                             
  at finite temperature.
  Spontaneous symmetry  breaking is related to 
 many phenomena in physics, such as,  superfluidity, superconductivity, ferro-magnetism,
  Bose-Einstein condensation etc. One well known example which will be studied here in some detail is 
 the chiral symmetry breaking in strong interactions. 
At low energies, the QCD vacuum is characterized by 
a non-vanishing expectation value
\cite{shifman}:
  $\langle \bar\psi\psi\rangle \approx$ (235 
MeV)$^3$,
the so called chiral condensate.
This quantity
describes the density of quark-antiquark pairs
found in the QCD vacuum and its non-vanishing value 
is a manifestation of chiral symmetry breaking
\cite{harris}.
The chiral symmetry is restored at finite temperature through a chiral phase transition
which is believed to be  first or second order depending on
the underlying global symmetry \cite{pisarski1}.

In the temperature range below the chiral transition point the 
thermodynamics  of quarks and gluons 
may be investigated using
 the linear sigma model \cite{gell} 
 which serves as an effective
model for the low-temperature
phase of QCD \cite{bilic,bilic1}.
The original sigma model is formulated as spontaneously broken
$\varphi^4$ theory with four real scalar fields
which constitute the ($\frac12$,$\frac12$) representation
of the chiral SU(2) $\times$ SU(2).
Hence, the model
falls in the O(4) universality class
owing to the isomorphism between the groups O(4) and SU(2) $\times$ SU(2).
We shall consider here a linear sigma model with 
spontaneously broken O($N$) symmetry, where $N\geq 2$. 
According to the Goldstone theorem,  the spontaneous symmetry
breaking yields massless particles called {\em Goldstone bosons}
the number of which depends on the rank of the remaining unbroken symmetry.
In the case of the O($N$) group in the symmetry broken phase,
i.e.,  
at temperatures below
the point of the  phase transition,
there will be $N-1$ Goldstone bosons 
which we will call the {\em pions}.
In the symmetry broken phase the pions, in spite of being massless,  propagate slower than light
owing to finite temperature effects
\cite{pisarski,son1,son2,bilic2}. Moreover, the pion velocity approaches zero at the
critical temperature.
In the following we will use the term ``chiral fluid'' to denote 
a hadronic fluid in the symmetry broken phase consisting
predominantly of massless pions. In our previous papers \cite{tolic,tolic2} we have demonstrated 
that perturbations in the chiral fluid undergoing a radial Bjorken expansion 
 propagate in curved geometry described by an effective analog metric of the Friedmann Robertson Walker (FRW) type
 with hyperbolic spatial geometry

As an application  of the AdS/CFT  duality  in 
terms of D7-brane embeddings \cite{babington} 
 the chiral phase transition can be regarded    
as a transition from the Minkowski to black hole embeddings of the D7-brane in a D3-brane background. 
This has been exploited by  Mateos, Myers, and Thomson \cite{mateos} who find a strong first order phase transition.
 Similarly, a first order chiral phase transition was found by Aharony, Sonnenschein, and 
Yankielowitcz \cite{aharony} and Parnachev and Sahakyan \cite{parnachev} in the Sakai-Sugimoto model
\cite{sakai}.
 In this paper we consider a model of a brane world universe
in which the chiral fluid lives on the 3+1 dimensional boundary of 
the AdS$_5$ bulk. We will  combine the linear sigma model with
 a  boost invariant  spherically symmetric  Bjorken type expansion \cite{lampert}
and  use  AdS/CFT techniques
to establish a relation between the effective analog geometry on the
boundary and the bulk geometry which satisfies the field equations with negative
cosmological constant. The formalism  presented here could
also  be applied to the calculation of two point functions, 
 Willson loops, and entanglement entropy for a spherically expanding Yang-Mills plasma 
 as it was recently done by Pedraza \cite{pedraza} for a linearly expanding ${\cal N}=4$ supersymmetric Yang-Mills
 plasma.

The remainder of the paper is organized as follows. 
In Sec.~\ref{expanding} we describe the dynamics of the chiral fluid and the
corresponding 3+1 dimensional analog geometry. 
In Sec.~\ref{holographic} we solve the Einstein equations in the bulk
using the metric ansatz that respects the spherical boost invariance of 
the fluid energy-momentum tensor at the boundary.
We demonstrate a relationship of our solution with the D3-brane solution of 
10 dimensional supergravity.
In Sec.~\ref{pion} 
we  establish a connection of the bulk geometry with
 the analog geometry on the boundary and derive 
 the temperature dependence of the pion velocity.
 In the concluding section,
Sec.~\ref{conclusion}, we summarize our results and
discuss  physical consequences.

\section{Expanding hadronic fluid}  
\label{expanding}

Consider a linear sigma model in a background medium at finite temperature
in a general curved spacetime. 
The dynamics of mesons in  such a medium is described by
an effective action with O($N$) symmetry \cite{bilic2}
\begin{equation}
S_{\rm eff} =
\int d^4x \sqrt{-G}\,
\left[\frac{1}{2}G^{\mu\nu}\partial_{\mu}\mbox{\boldmath$\Phi$}
\partial_{\nu}\mbox{\boldmath$\Phi$}
- \frac{c_\pi}{f^2}\left(\frac{m_0^2}{2}
 \mbox{\boldmath$\Phi$}^2
+ \frac{\lambda}{4}
 (\mbox{\boldmath$\Phi$}^2)^2
\right)\right],
\label{eq0101}
\end{equation}
where {\boldmath$\Phi$}  denotes a multicomponent scalar field
$\mbox{\boldmath$\Phi$} \equiv (\Phi_1, ...,\Phi_N)$. The effective metric tensor,
its inverse, and its determinant are 
\begin{equation}
G_{\mu\nu} =\frac{f}{c_{\pi}}
[g_{\mu\nu}-(1-c_{\pi}^2)u_{\mu}u_{\nu}] ,
\label{eq022}
\end{equation}
\begin{equation}
G^{\mu\nu} =
\frac{c_{\pi}}{f}
\left[g^{\mu\nu}-(1-\frac{1}{c_{\pi}^2})u^{\mu}u^{\nu},
\right],
\label{eq029}
\end{equation}
\begin{equation}
G \equiv \det G_{\mu\nu} = \frac{f^4}{c_{\pi}^2} \det g_{\mu\nu},
\label{eq030}
\end{equation}
respectively, where
$u^{\mu}$ is  the velocity of the fluid and
 $g^{\mu\nu}$ is the background metric.
 The coefficient
 $f$ and the pion velocity $c_\pi$ depend  on the local temperature $T$
 and on the parameters $\lambda$ and $m_0^2$ of the model,
 and may be calculated
in perturbation theory.
At zero temperature the medium is  absent in which case $f=c_\pi=1$
and $G_{\mu\nu}$ is identical to $g_{\mu\nu}$.

 If $m_0^{2} < 0$ the 
symmetry will be spontaneously broken.
At zero temperature the $\Phi_i$ fields develop
non-vanishing vacuum expectation values such that
\begin{equation}\label{eq2}
\sum_i \langle \Phi_i\rangle^{2} =
- \frac{m_0^{2}}{\lambda} \equiv f_{\pi}^{2} .
\end{equation}
Redefining the fields
$
\Phi_i(x) \rightarrow
 \langle \Phi_{i} \rangle+\varphi_i(x) ,
$
 the fields 
$\varphi_i$
 represent
quantum fluctuations around the vacuum expectation values
$\langle \Phi_i\rangle$.
It is convenient to choose here 
$\langle \Phi_{i} \rangle = 0$ for $i=1,2,..,N-1$, and $\langle \Phi_N \rangle =
f_{\pi}$ .
At nonzero temperature the quantity  
$\langle \Phi_N \rangle$, usually referred to as the condensate, is  temperature dependent
and vanishes at the point of phase transition.
In view of the usual physical meaning of the $\varphi$-fields in the chiral SU(2) $\times$ SU(2) sigma model 
it is customary 
to denote the $N-1$ dimensional vector $(\varphi_1, ...,\varphi_{N-1})$  by {\boldmath$\pi$},
the field $\varphi_N$ by $\sigma$, and the condensate $\langle \Phi_N\rangle$
by $\langle \sigma \rangle$.
In this way one obtains
 the effective Lagrangian
 in which
the O($N$) symmetry is explicitly broken down to O($N-1$).

At and above a critical temperature $T_{\rm c}$ the symmetry will be restored
and all the mesons will have the same mass.
Below the
critical temperature the meson masses 
are 
$m_{\pi}^2 =  0$,  and
$m_{\sigma}^2 = 2\lambda \langle\sigma\rangle^{2}$.  
The temperature dependence of $\langle \sigma \rangle$
is obtained by
minimizing the thermodynamic potential
with respect to
 $\langle \sigma \rangle$.
Given $N$, $f_\pi$, and $m_\sigma$, the extremum condition can be solved numerically at one loop order
\cite{bilic1}. In this way,
the value $T_{\rm c}= 183$ MeV of the critical temperature  was found \cite{tolic2}
for $N=4$, $f_\pi=92.4$ MeV, and $m_\sigma=1$ GeV as a phenomenological input.

Consider first a homogeneous steady flow of the medium
consisting of pions at finite temperature
in the Minkowski background with $g_{\mu\nu}=\eta_{\mu\nu}$.
In a comoving reference frame (characterized by $u_\mu=(1,0,0,0)$)
the effective metric (\ref{eq022}) is diagonal with corresponding 
line element 
\begin{equation}
ds^2= G_{\mu\nu}dx^\mu dx^\nu = fc_\pi dt^2-\frac{f}{c_\pi} \sum_{i=1}^{3} dx_i^2 .
\label{eq0243}
\end{equation}
At nonzero temperature, $f$ and $c_\pi$  can be derived from the
finite temperature self energy  $\Sigma(q,T)$ 
in the limit when the external momentum
$q$ approaches zero and
  can be expressed in terms of second derivatives of
 $\Sigma(q,T)$ with respect to $q_0$ and $q_i$.
The  quantities $f$ and $c_\pi$ as functions of temperature have been calculated
at one loop level by Pisarski and Tytgat \cite{pisarski}
in the low temperature approximation
\begin{equation}
 f\sim 1-\frac{T^2}{12f_\pi^2}-\frac{\pi^2}{9}\frac{T^4}{f_\pi^2 m_\sigma^2},
 \quad
 c_\pi\sim 1-\frac{4\pi^2}{45}\frac{T^4}{f_\pi^2 m_\sigma^2} ,
  \label{eq3024}
 \end{equation}
and by Son and Stephanov for temperatures
close to the chiral transition point \cite{son1,son2} 
(see also \cite{tolic}).
Whereas the low temperature result  (\ref{eq3024}) 
  does not depend on $N$,  the result near 
  the critical temperature does.
In the limit  $T\rightarrow T_{\rm c}$ in $d=3$ dimensions one finds \cite{son1,son2} 
\begin{equation}
f \propto (1-T/T_{\rm c})^{\nu -2\beta},
\quad  
c_\pi \propto (1-T/T_{\rm c})^{\nu/2} ,
\label{eq252}
\end{equation}
where 
the critical exponents $\nu$ and $\beta$ depend on $N$.
For example, $\nu=0.749$ and $\beta=0.388$ for the O(4) universality class \cite{hasenbusch,toldin}.
 Combining these limiting cases  with the numerical results at one loop order \cite{tolic2},
a reasonable fit 
in the entire range $0\leq T \leq T_{\rm c}$ 
is provided by
\begin{equation}
f =(1-T^4/T_{\rm c}^4)^{\nu -2\beta},
\quad c_\pi =(1-T^4/T_{\rm c}^4)^{\nu/2}.
\label{eq051}
\end{equation}
With this 
 we have $f=c_\pi=1$  at $T=0$,
 $c_\pi^2\simeq 1-\nu (T/T_{\rm c})^4$ near $T=0$ as predicted 
 by the one loop low temperature approximation \cite{pisarski}, and 
we recover the correct behavior 
(\ref{eq252})
near $T=T_{\rm c}$. 

Next we assume that the background medium is going through a  Bjorken type  expansion.
A realistic hydrodynamic  model of heavy ion collisions involves 
a transverse expansion
superimposed on a longitudinal boost invariant expansion.
Here we will consider
  a  radial boost invariant  Bjorken expansion \cite{bjorken}
in Minkowski background spacetime.
 A similar model has been previously studied in the context of disoriented
chiral condensate \cite{lampert}.
Our approach
is in spirit similar to that of Janik and Peschanski \cite{janik1,janik2}
who consider a hydrodynamic model based on
a longitudinal Bjorken expansion and 
 neglect the transverse expansion. 
A spherically symmetric Bjorken expansion considered here 
is certainly not the best model for description 
 of high energy heavy ion collisions but is
 phenomenologically relevant in the context of hadron production in $e^+e^-$.
 
  The Bjorken expansion is defined by a specific choice of the fluid 4-velocity which may be regarded as 
 a coordinate transformation in Minkowski spacetime. 
 In radial coordinates
$x^\mu=(t,r,\vartheta,\varphi)$  the fluid four-velocity 
of the radial Bjorken expansion is given by
\begin{equation}
u^\mu=(\gamma,\gamma v_r, 0,0)= (t/\tau, r/\tau,0,0),
\label{eq144}
\end{equation}
where $v_r=r/t $ is the radial three-velocity  and $\tau=\sqrt{t^2-r^2}$ is the 
{\em proper time}.
It is convenient to introduce the so called {\em radial rapidity} variable  $y$ 
and parameterize 
the four-velocity  as
\begin{equation}
u^\mu=(\cosh y,\sinh y,0, 0),
\label{eq146}
\end{equation}
so that
the radial three-velocity is
$v_r=\tanh y$.
Now, it is natural to use the spherical rapidity coordinates $(\tau, y)$  
defined by the following transformation 
\begin{equation}
t=\tau \cosh y ,
\quad
 r=\tau \sinh y .
\label{eq147}
\end{equation}
As in these coordinates the velocity components are  $u^\mu=(1,0,0,0)$,
the new coordinate frame is comoving.
The transformation from  $(t,r,\vartheta,\varphi)$ to 
$(\tau,y,\vartheta,\varphi)$  takes the background Minkowski metric into
\begin{equation}
g_{\mu\nu}={\rm diag} (1,  -\tau^2,-\tau^2\sinh^2\! y,-\tau^2\sinh^2\! y \sin^2 \vartheta ),
\label{eq218}
\end{equation}
which describes the geometry of the Milne cosmological model \cite{milne} -- a 
homogeneous, isotropic, expanding universe
with the cosmological scale  $a=\tau$ and negative spatial curvature.

The functional dependence of the fluid temperature $T$  on $\tau$ 
can be derived from
 energy-momentum conservation.
 First, we assume that our fluid is conformal, i.e., that its energy momentum is traceless $T^\mu_\mu=0$.
 Assuming quite generally that the pressure is not isotropic
 \begin{equation}
  T^\mu_\nu={\rm diag}(\rho, -p_y, -p_\bot, -p_\bot),
 \end{equation}
the tracelessness implies
$ \rho-p_y-2p_\bot=0$.
On the other hand, the energy momentum conservation 
${T^{\mu\nu}}_{;\nu}=0$
yields $p_\bot =p_y\equiv p$ which
implies that the fluid is 
perfect with energy-momentum tensor 
\begin{equation}
T_{\mu\nu}=(p+\rho) u_{\mu}u_{\nu}-p g_{\mu\nu}   ,
\label{eq001}
\end{equation}
Furthermore,  from the continuity equation
%
$u^\mu \rho_{;\mu}+(p+\rho){u^\mu}_{;\mu}=0$,
 one finds
\begin{equation}
\frac{\partial\rho}{\partial\tau}  + \frac{4\rho}{\tau}  =0 ,
\label{eq148}
\end{equation}
with solution
\begin{equation}
\rho=\left(\frac{c_0}{\tau}\right)^4.
\label{eq006}
\end{equation}
The dimensionless constant $c_0$  
 may be fixed from
the phenomenology of high energy collisions.
For example, with 
a typical value of
$\rho=1$  GeVfm$^{-3} \approx 5$ fm$^{-4}$ at  $\tau\approx 5$ fm 
\cite{kolb-russkikh}
we find
$c_0=7.5$.

Equation (\ref{eq006}) implies that  
the temperature of the expanding chiral fluid is, to a good approximation,  proportional to $\tau^{-1}$.
This follows from the fact that the expanding hadronic matter is dominated by massless pions,
and hence, the pressure of the fluid 
may be approximated by \cite{landau}
\begin{equation}
p=\frac13 \rho=\frac{\pi^2}{90}(N-1)T^4.
\label{eq106}
\end{equation}
This approximation is justified as long as we are not very close to $T=0$ 
in which case  we may neglect the contribution of the condensate
with vacuum energy equation of state $p=-\rho$.
Moreover, this approximation is consistent with the conformal fluid assumption
which also fails at and
near $T=0$,  because 
the energy momentum tensor of the vacuum is proportional to the metric tensor
and hence $T^\mu_\mu \neq 0$ in the vicinity of $T=0$.
Combining  (\ref{eq106}) with (\ref{eq006}) one finds
 \begin{equation}
 T =\left( \frac{30}{\pi^2(N-1)}\right)^{1/4}\frac{c_0}{\tau} .
\label{eq008}
\end{equation}
Hence, the temperature and proper time  are uniquely related.
For example, there is a unique proper time $\tau_{\rm c}$
which corresponds to 
 the critical temperature $T_{\rm c}$ of the chiral phase transition, so that
 \begin{equation}
 \frac{T}{T_{\rm c}} =\frac{\tau_{\rm c}}{\tau} .
\label{eq1108}
\end{equation}
 If we adopt the value $c_0=7.5$  estimated above  and  $T_{\rm c}=0.183$ GeV $=0.927$ fm$^{-1}$ \cite{tolic2}
 as the critical temperature for $N=4$,
the corresponding proper time will be 
$\tau_{\rm c}= 8.2 \; {\rm fm} =41\; {\rm GeV}^{-1}$.




The energy momentum tensor in comoving coordinates takes the form
\begin{equation}
T^{\rm conf}_{\mu\nu}
 =\frac{c_0^4}{3\tau^4}\,{\rm diag} (3, \tau^2,\tau^2\sinh^2\! y,\tau^2\sinh^2\! y \sin^2 \vartheta ).
\label{eq2008}
\end{equation}
 If we relax the conformal fluid condition $T^\mu_\mu = 0$ and add
the contribution of the vacuum 
to the conformal part, 
the  energy momentum tensor will read
\begin{equation}
T_{\mu\nu}
 =T_{\mu\nu}^{\rm conf}+\rho_{\rm vac} g_{\mu\nu} ,
\label{eq2108}
\end{equation}
where $\rho_{\rm vac}$ is a constant vacuum energy density. 
Then, instead of (\ref{eq148}), we obtain
\begin{equation}
\frac{\partial\rho}{\partial\tau}  + \frac{4(\rho-\rho_{\rm vac})}{\tau}  =0 ,
\label{eq2148}
\end{equation}
with solution $\rho=(c_0/\tau)^4+\rho_{\rm vac}$.
Combining this with the equation of state
corresponding to (\ref{eq2108})
we  obtain precisely the same relation (\ref{eq1108})
between  the temperature and proper time.

In the comoving coordinate frame defined by the coordinate transformation
(\ref{eq147})
the analog metric (\ref{eq022}) is diagonal with line element
\begin{equation}
ds^2= G_{\mu\nu}dx^\mu dx^\nu = f c_\pi d\tau^2-fc_\pi^{-1}\tau^2(dy^2+ \sinh^2\! y d\Omega^2),
\label{eq243}
\end{equation}
where $f$ and $c_\pi$ are 
given by (\ref{eq051}) and are functions of $\tau$ through 
(\ref{eq1108}).
 Hence, this metric represents  an FRW spacetime with negative
spatial curvature.

Note that
the spacetime described by (\ref{eq243}) with (\ref{eq252})
  has a curvature singularity at $\tau=\tau_{\rm c}$.
The Ricci scalar corresponding to (\ref{eq243}) is given by
   \begin{equation}
 R=  \frac{3}{fc_\pi }\left[\frac{2(1-c_\pi^2)}{\tau^2}
 +
 \frac{3\dot{f}}{\tau f}-\frac{5\dot{c}_\pi}{\tau c_\pi}
 -\frac{1}{2}
 \frac{\dot{f}^2}{f^2}+\frac{5}{2}\frac{\dot{c}_\pi^2}{c_\pi^2}
 +
 \frac{\ddot{f}}{f}-\frac{\ddot{c}_\pi}{c_\pi}
 -\frac{2\dot{f}\dot{c}_\pi}{f c_\pi}\right],
 \label{eq2051}
 \end{equation} 
  where the overdot denotes a derivative with respect to $\tau$.
 Using (\ref{eq252}) in the limit $\tau\rightarrow \tau_{\rm c}$ one finds that   $R$ 
 diverges as
   \begin{equation}
 R \sim(\tau-\tau_{\rm c})^{2\beta-3\nu/2-2}.
 \end{equation}

\section{Holographic description of the hadronic fluid}
\label{holographic}
We now turn to the AdS/CFT correspondence and look for a five-dimensional  bulk geometry
dual to the four-dimensional spherically expanding chiral fluid described by the energy momentum tensor
(\ref{eq001}). A general asymptotically AdS metric in Fefferman-Graham coordinates \cite{fefferman}
is of the form
\begin{equation}
ds^2=g_{AB}dx^Adx^B =\frac{\ell^2}{z^2}\left( h_{\mu\nu} dx^\mu dx^\nu -dz^2\right),
 \label{eq3001}
\end{equation}
where we use the uppercase Latin alphabet for bulk indices
and the Greek alphabet for 3+1 spacetime indices.
Our curvature conventions are as follows:
$R^{a}{}_{bcd} = \partial_c \Gamma_{db}^a - 
\partial_d \Gamma_{cb}^a + \Gamma_{db}^e \Gamma_{ce}^a  - \Gamma_{cb}^e \Gamma_{de}^a$ and $R_{ab} = R^s{}_{asb}$, 
so that Einstein's equations are $R_{ab} - \frac{1}{2}R g_{ab} = +\kappa T_{ab}.$

The length scale $\ell$ is the AdS curvature radius
related to the  cosmological constant by
$\Lambda= -6/\ell^2$.
The four dimensional tensor $h_{\mu\nu}$ may be  expanded near the boundary at $z=0$ 
as \cite{haro}
\begin{equation}
 h_{\mu\nu}=g^{(0)}_{\mu\nu}+z^2 g^{(2)}_{\mu\nu}+z^4 g^{(4)}_{\mu\nu}+z^6
g^{(6)}_{\mu\nu}+\ldots \ ,
\label{eq3002}
\end{equation}
where $g^{(0)}_{\mu\nu}$ is the background metric on the boundary.

Let us assume now that the boundary geometry is described by
the Ricci flat spacetime metric (\ref{eq218}). 
According to the holographic renormalization rules \cite{haro} in this case $g^{(2)}_{\mu\nu}=0$ 
and $g^{(4)}_{\mu\nu}$ is proportional to the vacuum
expectation value of the energy-momentum tensor 
\begin{equation}
 g^{(4)}_{\mu\nu}=-\frac{4\pi G_5}{\ell^3} \langle T^{\rm conf}_{\mu\nu}\rangle   ,
 \label{eq3003}
\end{equation}
where $G_5$ is the five dimensional Newton constant and 
the expectation value on the righthand side is assumed to be equal to the energy momentum tensor
(\ref{eq2008}).
This equation is an explicit realization of the AdS/CFT prescription
that the field dual to the energy momentum tensor $T_{\mu\nu}$ should be
the four-dimensional metric $g_{\mu\nu}$.

Instead of a linear boost invariance of \cite{janik1,janik2,janik3},
we impose a spherically symmetric boost invariance.
The most general metric respecting the spherically symmetric boost invariance
in Fefferman-Graham coordinates is of the form
\begin{equation}
ds^{2} = \frac{\ell^2}{z^2}
\left[A(z, \tau)d\tau^2 - \tau^2 B(z, \tau)(dy^2 + \sinh^2\! y d\Omega^2) - dz^2\right].
\label{eq3004}
\end{equation}
Equations (\ref{eq3002}) and (\ref{eq3003}) together with
(\ref{eq2008})
imply the conditions near the $z=0$ boundary
 \begin{equation}
A=1-3 k z^4/\tau^4 +\ldots \ , \quad B=1+k z^4/\tau^4 +\ldots \ , 
\label{eq3005}
\end{equation}
where 
\begin{equation}
k= \frac{4\pi G_5 c_0^4}{3\ell^3}
 \label{eq3105}
\end{equation}
is a dimensionless constant.

Using the metric ansatz (\ref{eq3004}) we
solve  Einstein's equations with negative cosmological constant
\begin{equation}
R_{AB}-\left(\frac{1}{2}R-\frac{6}{\ell^2}\right)g_{AB}=0.
\label{eq3006}
\end{equation}
By inspecting the components of (\ref{eq3006}) subject to (\ref{eq3004}), 
it may be verified  that Einstein's equations are invariant under simultaneous rescaling
$\tau\rightarrow \lambda \tau$ and $z\rightarrow \lambda z$,
for any real positive $\lambda$.
In other words, if $A=A(z,\tau)$ and 
$B=B(z,\tau)$ are solutions to (\ref{eq3006}), then so are 
$A=A(\lambda z,\lambda \tau)$ and $B=B(\lambda z,\lambda\tau)$.
This implies that
 $A$ and $B$ are functions of 
a single scaling variable
$v = z/\tau$.
From the $zz$ and $z\tau$ components of Einstein's equations we find
two independent differential equations for $A$ and $B$
\begin{equation}
\frac{2{A'}}{v A} + \frac{{A'}^2}{A^2} + \frac{6{B'}}{vB} +
\frac{3{B'}^2}{B^2} - \frac{2{A''}}{A} - \frac{6{B''}}{B}=0,
\label{eq3009}
\end{equation}
\begin{equation}
\frac{6{A'}}{A} - \frac{v{A'}{B'}}{AB} - \frac{3v{B'}^2}{B^2} + \frac{6v{B''}}{B}=0,
\label{eq3010}
\end{equation}
where the prime denotes a derivative with respect to $v$.
It may be easily verified that the functions
\begin{equation}
A(v)=\frac{(1-kv^4)^2}{1+kv^4}, \quad
B(v)=1+kv^4.
\label{eq3012}
\end{equation}
satisfy (\ref{eq3009}), (\ref{eq3010}), 
and the remaining set of Einstein's equations, $k$ being an arbitrary constant.
Equations (\ref{eq3012}) satisfy the boundary conditions (\ref{eq3005})
if we identify $k$ with the constant defined in (\ref{eq3105}).
The line element (\ref{eq3004}) becomes
\begin{equation}
ds^{2} = \frac{\ell^2}{z^2}
\left[ \frac{(1-k z^4/\tau^4)^2}{1+k z^4/\tau^4} d\tau^2 -  (1+k z^4/\tau^4)\tau^2 (dy^2 + \sinh^2\! y d\Omega^2) - dz^2\right].
\label{eq3013}
\end{equation}
This type of metric is a special case of a more general solution derived by 
Apostolopoulos, Siopsis, and Tetradis \cite{apostolopoulos,tetradis}
with an arbitrary FRW cosmology at the boundary.

It is useful to compare the solution (\ref{eq3013}) with the
static
Schwarzschild-AdS$_5$ metric \cite{chamblin}
\begin{equation}
ds^{2} = \frac{\ell^2}{z^2}
\left[ \frac{(1-z^4/z_0^4)^2}{1-\kappa z^2/(2\ell^2)+z^4/z_0^4} d\tau^2 - 
(1-\kappa z^2/(2\ell^2)+z^4/z_0^4)\ell^2 d\Omega_3^2(\kappa) - dz^2\right] ,
\label{eq3216}
\end{equation} 
where $\kappa=0,1,-1$ for a flat, spherical and hyperbolic boundary  geometry with
\begin{equation}
d\Omega_3^2(\kappa)
 =
\left\{\begin{array}{ll}
  dy^2+\sinh^2 y d\Omega^2,     & \quad \kappa=-1 ,  \\
dy^2 +y^2d\Omega^2,        & \quad  \kappa=0 , \\
dy^2+\sin^2 y d\Omega^2,         &\quad \kappa=1  .
\end{array} \right. 
\label{eq2208}
\end{equation}
The location of the horizon
 $z_0$ is related to the BH mass as
\begin{equation}
z_0^4=\frac{16\ell^4}{4\mu+\kappa^2} ,
\label{eq2209}
\end{equation}  
where $\mu$ is the BH mass in units of $\ell^{-1}$.
It has been noted \cite{kajantie} that (\ref{eq3013}) is obtained from 
(\ref{eq3216}) by keeping the conformal factor $\ell^2/z^2$ 
and elsewhere making the replacements $\kappa \rightarrow \kappa+1$ 
and $\ell \rightarrow \tau$.
Hence, the constant $k$ in (\ref{eq3013}) is related to the BH mass of the static solution as
$k=\mu/4$.

It is worth emphasizing the difference between the spherically symmetric solution (\ref{eq3013})
and the solution of the similar form found in the case of linear boost invariance \cite{janik1}.
First, our solution (\ref{eq3013}) is {\em exact} and valid at all times.
In contrast,  
the  solution found in \cite{janik1} of the  form (\ref{eq3012}) with $v \sim z/\tau^{1/3}$
   is valid only  
in the asymptotic regime $\tau\rightarrow \infty$.
A similar late time asymptotic solution was found  
by Sin, Nakamura, and Kim \cite{sin}
for the case of a linear 
anisotropic expansion described by the Kasner metric.
In a related recent work 
Fischetti, Kastor, and Traschen \cite{fischetti} have 
 constructed solutions that expand spherically and approach the Milne universe at late times.
 Their solution, obtained by making use of a special type of ideal fluid in 
addition to the negative cosmological constant in the bulk,
gives rise to open FRW cosmologies at the boundary and on the Poincar\'e slices
(which correspond to the z-slices in Fefferman Graham coordinates) of
a late time asymptotic AdS$_5$.

Another remarkable property of the solution  (\ref{eq3013}) is that 
the induced metric on each $z$-slice is equivalent to the Milne metric.
This may be seen as follows. 
The  3+1 dimensional metric induced on a $z$-slice is, up to a multiplicative constant,
 given by
\begin{equation}
ds^{2} = 
 \frac{(1-k z^4/\tau^4)^2}{1+k z^4/\tau^4}d\tau^2 -  (1+k z^4/\tau^4)\tau^2 (dy^2 + \sinh^2\! y d\Omega^2).
\label{eq3113}
\end{equation}
This line element is of the form (\ref{eq243}) and the corresponding 3+1 spacetime
at a given $z$-slice
may be regarded as 
an FRW spacetime.
Then the coordinate transformation  
$\tilde{\tau}(\tau)=\tau (1+k z^4/\tau^4)^{1/2}$
brings the metric (\ref{eq3113}) to the Milne form (\ref{eq218}).

The solution (\ref{eq3013}) is closely related 
to the D3-brane solution of 10 dimensional supergravity 
corresponding to a stack of $N_{\rm D}$ coincident D3-branes.
A  near-horizon nonextremal D3-brane metric is given by  \cite{horowitz}
\begin{equation}
ds^{2} = \frac{U^2}{L^2}
\left[ \left(1-\frac{U_0^4}{U^4}\right) dt^2  - \frac{L^4}{U^4}\left(1-\frac{U_0^4}{U^4}\right)^{-1}dU^2 
-\sum^3_{i=1} dy_i^2\right]-L^2 d\Omega_5^2 ,
\label{eq3017}
\end{equation}
where
$L^2=\ell_s^2\sqrt{4\pi g_sN_{\rm D}}$,
$g_s$ is 
the string coupling constant, and 
 $\ell_s=\sqrt{\alpha'}$ is the fundamental string length.
 Ignoring the five sphere which decouples throughout the spacetime (\ref{eq3017}),
 the remaining five dimensional spacetime  is equivalent to
 the standard AdS$_5$ Schwarzschild  spacetime in the limit of large BH mass \cite{witten2}
 and is asymptotically AdS$_5$.
 By identifying $L$ with 
 the AdS curvature radius $\ell $, 
  replacing the constant $U_0$ with $U_0=z_0/\sqrt{2}$,
 rescaling
the coordinates 
\begin{equation}
t=\frac{2\ell^2}{z_0^2}\tau, \quad 
y_i= \frac{2\ell^2}{z_0^2} x_i,  
 \label{eq3019}
\end{equation}
and making a coordinate transformation $U\rightarrow z$
\begin{equation}
U=\frac{z_0^2}{2z}\sqrt{1+\frac{z^4}{z_0^4}} ,
 \label{eq3119}
\end{equation}
the metric of the asymptotically AdS$_5$ bulk of (\ref{eq3017}) turns into
\begin{equation}
ds^{2} = \frac{\ell^2}{z^2}
\left[ \frac{(1-z^4/z_0^4)^2}{1+z^4/z_0^4} d\tau^2 -  (1+z^4/z_0^4)\sum^3_{i=1} dx_i^2 - dz^2\right].
\label{eq3016}
\end{equation} 
This coincides with equation (\ref{eq3216}) for $\kappa=0$ with $z_0$ related to the BH mass as $z_0^4=4\ell^4/\mu$.
In this coordinate representation, the BH horizon is at $z=z_0$
and the inverse horizon temperature is
 $\beta=\pi z_0/\sqrt{2}$.

 In the case $\kappa=+1$ or $-1$ the metric (\ref{eq3016}) may be regarded as a large BH mass limit of a 
Schwarzschild-AdS$_5$ metric given by (\ref{eq3216}) with (\ref{eq2208}).
In the limit $\mu\rightarrow \infty$ we have $z_0/\ell \rightarrow 0$,
so  $z^2/l^2\ll 1$ for $z$ close to the horizon and the quadratic term in the metric coefficients 
in (\ref{eq3216}) vanishes in that limit.
Hence, taking $\mu\rightarrow \infty$ in (\ref{eq3216}) for $\kappa=+1$ or $-1$  one  finds
\begin{equation}
ds^{2} = \frac{\ell^2}{z^2}
\left[ \frac{(1-z^4/z_0^4)^2}{1+z^4/z_0^4} d\tau^2 -  (1+z^4/z_0^4)\ell^2 d\Omega_3^2(\kappa) - dz^2\right].
\label{eq3316}
\end{equation}
Comparing with this, the geometry  (\ref{eq3013}) appears as a dynamical black hole
with the location of the horizon $z_0=\tau/k^{1/4}$ moving in the bulk with velocity $k^{-1/4}$. 
Then the horizon temperature depends on time as
\begin{equation}
 T=\frac{\sqrt{2}k^{1/4}}{\pi\tau} ,
 \label{eq3020}
 \end{equation}
 in agreement with Tetradis \cite{tetradis}.

 \begin{figure}[t]
\begin{center}
\includegraphics[width=0.8\textwidth,trim= 0 0cm 0 0cm]{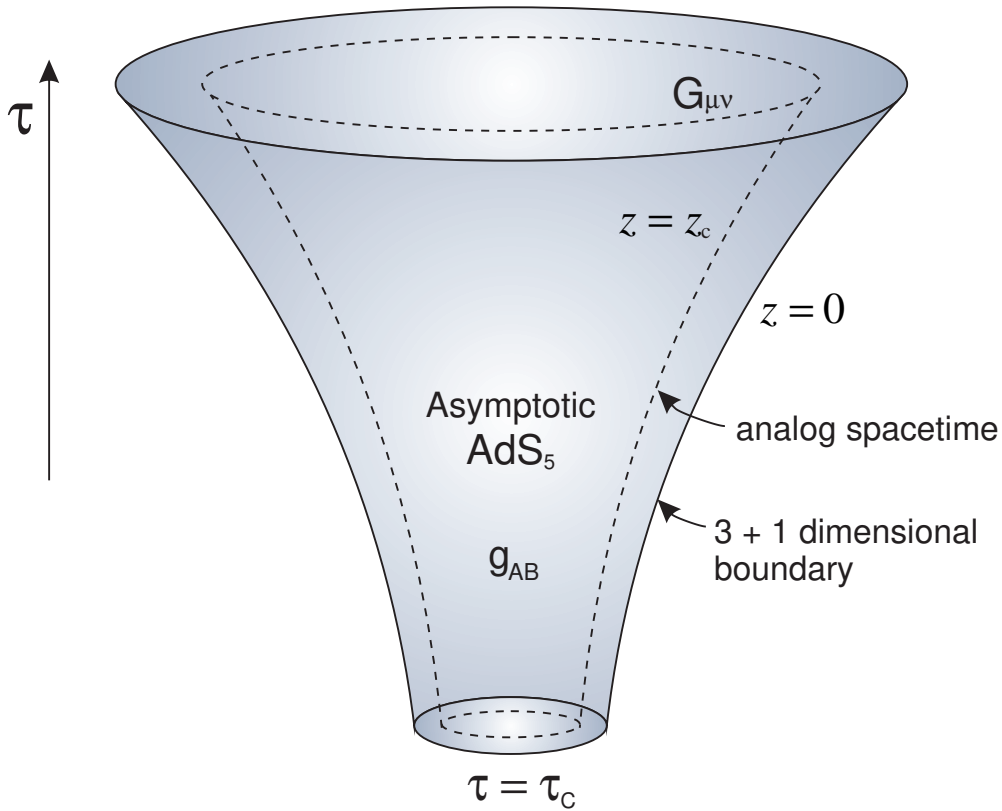}
\caption{An illustration of the correspondence between the asymptotic AdS geometry 
in the bulk with the analog spacetime geometry on its $3+1$ boundary. 
The induced metric  on the $z_c$-slice  corresponds to 
the effective metric (\ref{eq243}) in the symmetry broken phase ($\tau>\tau_c)$.
}
\label{fig1}
\end{center}
\end{figure}

 \section{The pion velocity}
 \label{pion}

In this section we  exploit the finite temperature AdS/CFT correspondence.
We first consider the dynamical case of an expanding fluid which  
is phenomenologically relevant to high energy collisions
and make use of  the relation between the bulk asymptotic AdS geometry (\ref{eq3013})
and the Bjorken dynamics at the boundary.
The correspondence 
is established by making use of the following assumptions:
\begin{enumerate}
\item[a)]
The horizon temperature (\ref{eq3020}) is proportional to the physical temperature of the 
expanding conformal fluid.
\item[b)]
There exist a maximal $z$ equal to $z_{\rm c}=k^{-1/4}\tau_{\rm c}$   where the critical proper time  $\tau_{\rm c}$ 
corresponds to the critical temperature $T_{\rm c}$. 
 \item[c)]  
 The induced metric (\ref{eq3113}) on the $z_c$-slice  corresponds to 
the effective metric (\ref{eq243}) in the symmetry broken phase ($\tau>\tau_c)$
in which the perturbations (massless pions) 
propagate.
\end{enumerate}
The first assumption stems from the relation 
(\ref{eq008}) and is obviously in agreement with the Bjorken dynamics.
The assumption b) is similar to that of Erlich et al.\ \cite{erlich2}
who assumed the infrared cutoff at some $z=z_m$ (``infrared brane'').
Our key assumption c) is motivated by the apparent resemblance of the 
effective analog metric (\ref{eq243})  to the  induced metric (\ref{eq3113}). 
The geometry is illustrated  in  Fig.\ \ref{fig1}.
The comparison of the  induced metric (\ref{eq3113}) with the
effective analog metric (\ref{eq243}) (with  (\ref{eq051}))  
 yields
\begin{equation}
 f=1-\tau_{\rm c}^4/\tau^4, \quad\quad c_\pi=\frac{1-\tau_{\rm c}^4/\tau^4 }{1+\tau_{\rm c}^4/\tau^4 }.
 \label{eq3122}
 \end{equation}

 A similar correspondence may be drawn by considering
the static case and relate the properties of the chiral fluid 
to the Schwarzschild-AdS black hole. We make use of slightly modified assumptions a)-b):
\begin{enumerate}
\item[a')]
A correspondence exists between the fifth coordinate $z$ and the physical temperature $T$ of the chiral fluid 
such that  $z$ is proportional to $1/T$.
\item[b')]
The horizon temperature defined in  (\ref{eq3020}) is proportional to the critical temperature of the 
chiral phase transition.
\item[c')]  
 The metric  induced on a  $z$-slice from the bulk metric (\ref{eq3016}) corresponds to 
the effective metric (\ref{eq0243}) in the symmetry broken phase ($T<T_{\rm c})$.
\end{enumerate}
According to a') and b') 
we identify  $z/z_0 \equiv T_{\rm c}/T$ and comparing the effective metric (\ref{eq0243}) with (\ref{eq3016})
we find
 \begin{equation}
  f=1-T^4/T_{\rm c}^4, \quad\quad c_\pi=\frac{1-T^4/T_{\rm c}^4}{1+T^4/T_{\rm c}^4}.
 \label{eq3021}
 \end{equation}
 These equations are equivalent to  (\ref{eq3122})
 owing to the relation  (\ref{eq1108}) between the  temperature and proper time
 which is a consequence of energy conservation in the Bjorken dynamics.
 
\begin{figure}[t]
\begin{center}
\includegraphics[width=0.8\textwidth,trim= 0 0cm 0 0cm]{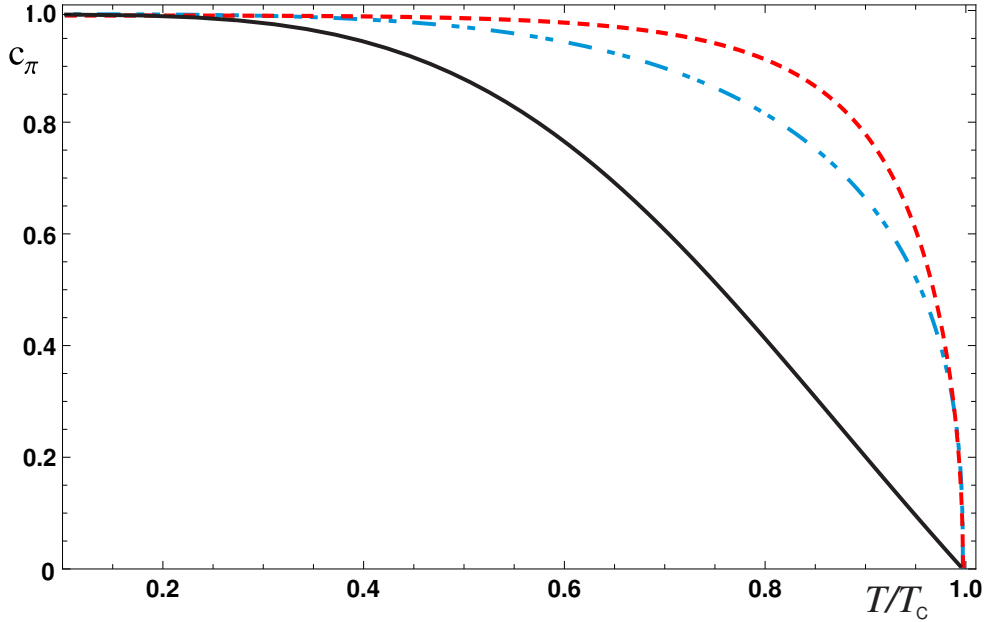}
\caption{Pion velocity versus $v\equiv T/T_{\rm c}$
described by three different functions:
approximate model $(1-v^4)^{\nu/2}$ based on the O(4) critical exponent $\nu = 0.749$ (dot-dashed blue);
the model based on numerical results  at one loop level \cite{tolic2} (dashed red);
the AdS/CFT model (\ref{eq3021}) (full black line).
}
\label{fig2}
\end{center}
\end{figure}

The  expression for the pion velocity in (\ref{eq3021}) (or in (\ref{eq3122}))
gives a roughly correct overall behavior in the temperature interval $(0,T_c)$
(Fig.\ \ref{fig2}). 
It is worth analyzing our predictions in the limiting cases of temperatures
near the endpoints of this interval.

In the limit $T\rightarrow 0$ 
the pion velocity in (\ref{eq3021})
 will agree with the low temperature approximation 
  (\ref{eq3024})
 if we identify 
 \begin{equation}
 T_{\rm c}=\left(\frac{45}{2\pi^2}f_\pi^2 m_\sigma^2\right)^{1/4}.
  \label{eq3025}
 \end{equation}
Our result 
confirms the expectation \cite{pisarski,pisarski2} that the deviation of the velocity squared 
from unity is proportional to
the free energy density, or pressure which for massless pions is given by (\ref{eq106}). 
Given $f_\pi$ and $m_\sigma$, equation  (\ref{eq3025}) 
can be regarded as  a prediction for the critical temperature. 
 The Particle Data Group \cite{beringer} gives a rather wide range 
 400-1500 MeV  of the sigma meson masses.
 With the lowest value $m_\sigma \simeq$ 400 MeV and $f_\pi=92.4$ MeV one finds 
 the lower bound 
 $T_{\rm c}\simeq$ 230 MeV which is somewhat larger than lattice results 
 which range between 150 and 190 MeV.
 
  It is important to note here that we do not recover the quadratic term 
 in the low temperature approximation (\ref{eq3024}) of the function $f$. 
 The reason may be that by assuming exact conformal invariance, i.e., the condition
 $T_\mu^\mu=0$, 
 we discarded the contribution of the vacuum energy 
(including the condensate) which actually dominates at  low temperatures
or equivalently at late times.

 As to the limit $T\rightarrow T_{\rm c}$,  
the behavior of our solution 
 differs in two aspects from what one finds in other treatments based on conventional calculations. 
 First, the induced metric (\ref{eq3113}) being equivalent to the Milne metric is Ricci flat so the
singularity at $\tau=z$  is just a coordinate singularity.
In contrast, as we have mentioned at the end of Sec.\ \ref{expanding},
the analog metric (\ref{eq243}) obtained from the linear sigma model exhibits  a curvature singularity
at the critical point $\tau=\tau_c$. 
Second, we do not recover
the critical exponents predicted by conventional calculations.
It is clear that the critical behavior  
differs significantly from the prediction based on
the O(4) critical exponents or
the one loop sigma model prediction. 
In the vicinity of the critical point  the function (\ref{eq3021})  approaches zero as  
\begin{equation}
 c_\pi \sim T_{\rm c}-T.
 \label{eq3022}
 \end{equation}
In contrast, the sigma model at one loop order \cite{tolic2} gives $c_\pi \sim  (T_{\rm c}-T)^{1/4}$ ,
whereas the Monte Carlo calculations  of the critical exponents for the O(4) universality class
 \cite{hasenbusch} yields $c_\pi \sim  (T_{\rm c}-T)^{0.37}$.

As a side remark, our bulk spacetime is free of curvature singularities.
Clearly, the Ricci scalar $R=20$ is regular everywhere.
However,
as noted in \cite{janik1},
there is a potential singularity of  
the Riemann tensor squared
$\Re^2\equiv R^{\mu\nu\rho\sigma} R_{\mu\nu\rho\sigma}$ 
at the hypersurface $z=\tau$.
A straightforward calculation yields 
\begin{equation}
\Re^2 =8\left(5+ \frac{144k^2v^8}{(1+kv^4)^4}    \right) ,
 \label{eq3023}
\end{equation}
which is regular everywhere.
Remarkably, if one substitutes  $w^4=3 z^4/\tau^{4/3}$ for
 our $kv^4=kz^4/\tau^4$ in  (\ref{eq3023}) the resulting expression for $\Re^2$ will be precisely
 of the  form obtained  in the asymptotic 
regime $\tau\rightarrow \infty$ \cite{janik1}
for the case of a perfect fluid undergoing a longitudinal Bjorken expansion.

\section{Conclusions}
\label{conclusion}

We have investigated a spherically expanding hadronic fluid
in the framework of AdS$_5$/CFT correspondence.
According to the holographic renormalization, the  energy momentum tensor of the 
spherically expanding conformal fluid  is related to
the bulk geometry described by the metric (\ref{eq3013})
which satisfies the field equations with negative
cosmological constant. 
It is remarkable that the exact correspondence exists  at all times $0\leq \tau <\infty$.
Based on this solution and analogy with the AdS-Schwarzschild black hole,
we have established a relation between the effective analog geometry on the
boundary and the bulk geometry. 
Assuming that the chiral fluid dynamics at finite temperature is described by
the linear sigma model as the underlying field theory,
we obtain a prediction for the pion velocity in the range
of temperatures below the phase transition point.
Compared with the existing conventional
calculations, a reasonable agreement is achieved generally
for those quantities, such as the pion velocity and the critical temperature,
that do not substantially depend on the number of scalars $N$.
In particular, our prediction at low temperature  confirms
the expectation \cite{pisarski,pisarski2} that the deviation of the pion velocity  
from the velocity of light is proportional to
the free energy density. 
The agreement  is of course not so good  for the critical exponents
since their values crucially depend on $N$. 
The estimate of the critical temperature is close to but somewhat higher then
the lattice QCD prediction. 

Obviously, our results are based on a crude simplification that the hadronic fluid 
is a  perfect conformal fluid undergoing a spherically symmetric radial expansion.
 A realistic hadronic fluid is neither perfect nor conformal.
 First, a hadronic fluid in general has a non vanishing shear viscosity 
 which is neglected here.
Second, our model is based on a scalar field theory with broken symmetry
which is only approximately conformal in the vicinity of the critical point where
the condensate vanishes and all particles (mesons and quarks) become massless.
Hence, in this way we could not have obtained more then a rough
 estimate of the critical temperature and the pion velocity at finite temperature.

\subsection*{Acknowledgments}
This work  was supported by the Ministry of Science,
Education and Sport
of the Republic of Croatia 
and the work of N.B. and D.T. was
partially supported by the ICTP-SEENET-MTP grant PRJ-09 ``Strings and Cosmology`` 
in the frame of the SEENET-MTP Network.


\begin{thebibliography}{99}

\bibitem{maldacena}
J.M.~Maldacena, 
Adv.\ Theor.\ Math.\ Phys.\ {\bf2}, 231 (1998); Int.\ J.\ Theor.\ Phys.\ {\bf 38}, 1113 (1999)
 [arXiv:hepth/9711200].
 
 \bibitem{gubser}
S.S.~ Gubser, I.R.~Klebanov, and A.M.~Polyakov, 
Phys. Lett. B {\bf428}, 105 (1998) [arXiv:hep-th/9802109].


\bibitem{witten1}
E.~Witten, 
Adv.\ Theor.\ Math.\ Phys.\ {\bf 2}, 253 (1998)
[arXiv:hep-th/9802150].


  \bibitem{rangamani} 
  M.~Rangamani,
  Class.\ Quant.\ Grav.\  {\bf 26}, 224003 (2009)
  [arXiv:0905.4352 [hep-th]].
  
  \bibitem{kovtun} 
  P.~Kovtun, D.~T.~Son and A.~O.~Starinets,
  Phys.\ Rev.\ Lett.\  {\bf 94}, 111601 (2005)
  [hep-th/0405231].
  
  \bibitem{erlich} 
  J.~Erlich,
  Int.\ J.\ Mod.\ Phys.\ A {\bf 25}, 411 (2010)
  [arXiv:0908.0312 [hep-ph]].
  
\bibitem{shifman} 
M.A.~Shifman,   Ann.\ Rev.\ Nucl.\ Part.\ Sci.\ {\bf 33} 199 (1983).  


\bibitem{harris}
    J. W. Harris and B. M\"uller,
    Ann.\ Rev.\ Nucl.\ Part.\ Sci.\ {\bf 46} 71 (1996).
    
\bibitem{pisarski1}
    R.\ D.\ Pisarski and F.\ Wilczek,
    Phys.\ Rev.\ D {\bf 29}, 338 (1984).  
            
  \bibitem{gell}
    M.\ Gell-Mann and M.\ L\'{e}vy,
    Nuovo Cimento {\bf 16} 705 (1960).
    
    \bibitem{bilic}
    N.\ Bili\'c, J.\ Cleymans, and M.\ D.\ Scadron,
    Int.\ J.\ Mod.\ Phys.\ {\bf A10} 1169 (1995).
%
\bibitem{bilic1}
    N.\ Bili\'c and H.\ Nikoli\'c,
    Eur.\ Phys.\ J.\ C {\bf 6}, 515 (1999).
    
     \bibitem{pisarski}
    R.\ D.\ Pisarski and M.\ Tytgat,
    Phys.\ Rev.\ D {\bf 54}, R2989 (1996);
    
\bibitem{son1}
    D.\ T.\ Son  and M.\ A.\ Stephanov,
    Phys.\ Rev.\ Lett.\ {\bf 88}, 202302 (2002).
    
\bibitem{son2}
    D.\ T.\ Son  and M.\ A.\ Stephanov,
    Phys.\ Rev.\ D {\bf 66}, 076011 (2002).
    
\bibitem{bilic2} 
  N.~Bili\'c and H.~Nikoli\'c,
  Phys.\ Rev.\ D {\bf 68}, 085008 (2003); hep-ph/0301275.
  
  \bibitem{tolic} 
  N.~Bili\'c and D.~Toli\'c,
  Phys.\ Lett.\ B {\bf 718}, 223 (2012)
  [arXiv:1207.2869 [hep-th]];
\bibitem{tolic2} 
 N.~Bili\'c and D.~Toli\'c,
  Phys.\ Rev.\ D {\bf 87}, 044033 (2013)
  [arXiv:1210.3824 [gr-qc]];
 
    \bibitem{babington} 
  J.~Babington, J.~Erdmenger, N.~J.~Evans, Z.~Guralnik and I.~Kirsch,
  Phys.\ Rev.\ D {\bf 69}, 066007 (2004)
  [hep-th/0306018].
  
\bibitem{mateos}   
  D.~Mateos, R.~C.~Myers and R.~M.~Thomson,
  Phys.\ Rev.\ Lett.\  {\bf 97}, 091601 (2006)
  [hep-th/0605046]. 
  
\bibitem{aharony} 
  O.~Aharony, J.~Sonnenschein and S.~Yankielowicz,
  Annals Phys.\  {\bf 322}, 1420 (2007)
  [hep-th/0604161]. 
 
\bibitem{parnachev} 
  A.~Parnachev and D.~A.~Sahakyan,
  Phys.\ Rev.\ Lett.\  {\bf 97}, 111601 (2006)
  [hep-th/0604173]. 
  
  \bibitem{sakai} 
  T.~Sakai and S.~Sugimoto,
  Prog.\ Theor.\ Phys.\  {\bf 113}, 843 (2005)
  [hep-th/0412141].
  
 \bibitem{lampert} 
  M.~A.~Lampert, J.~F.~Dawson and F.~Cooper,
  Phys.\ Rev.\ D {\bf 54}, 2213 (1996);
  G.~Amelino-Camelia, J.~D.~Bjorken and S.~E.~Larsson,
  Phys.\ Rev.\ D {\bf 56}, 6942 (1997);
  M.~A.~Lampert and C.~Molina-Paris,
  Phys.\ Rev.\ D {\bf 57}, 83 (1998);
  A.~Krzywicki and J.~Serreau,
   Phys.\ Lett.\ B {\bf 448}, 257 (1999).
 \bibitem{pedraza} 
  J.~F.~Pedraza,
  Phys.\ Rev.\ D {\bf 90}, 046010 (2014)
  [arXiv:1405.1724 [hep-th]].
  
 \bibitem{hasenbusch}
    M.~Hasenbusch, J.\ Phys.\ A {\bf 34} (2001) 8221
   [cond-mat/0010463]
   
\bibitem{toldin}
  F.~Parisen Toldin, A.~Pelissetto and E.~Vicari,
  JHEP {\bf 0307}, 029 (2003)
  [hep-ph/0305264].
     
\bibitem{bjorken} 
  J.~D.~Bjorken,
  Phys.\ Rev.\ D {\bf 27}, 140 (1983).

\bibitem{janik1} 
  R.~A.~Janik and R.~B.~Peschanski,
  Phys.\ Rev.\ D {\bf 73}, 045013 (2006)
  [hep-th/0512162]. 
  \bibitem{janik2} 
  R.~A.~Janik and R.~B.~Peschanski,
  Phys.\ Rev.\ D {\bf 74}, 046007 (2006)
  [hep-th/0606149].
  
   \bibitem{milne}
E.A.~Milne, Nature 130 (1932) 9.

  \bibitem{kolb-russkikh} 
  P.~F.~Kolb, J.~Sollfrank, and U.~W.~Heinz,
  Phys.\ Rev.\ C {\bf 62}, 054909 (2000).

\bibitem{landau}
 L.D.~Landau and E.M.~Lifshitz,
 {\em Statistical Physics},
(Pergamon, Oxford, 1993) p.\ 187.

\bibitem{fefferman}
C.~Fefferman and C.~R.~Graham,
 	arXiv:0710.0919 [math.DG]
 	
\bibitem{haro} 
  S.~de Haro, S.~N.~Solodukhin and K.~Skenderis,
  Commun.\ Math.\ Phys.\  {\bf 217}, 595 (2001)
  [hep-th/0002230]. 
  
 \bibitem{janik3} 
  R.~A.~Janik,
  Phys.\ Rev.\ Lett.\  {\bf 98}, 022302 (2007)
  [hep-th/0610144]. 
  
   \bibitem{apostolopoulos} 
  P.~S.~Apostolopoulos, G.~Siopsis and N.~Tetradis,
  Phys.\ Rev.\ Lett.\  {\bf 102}, 151301 (2009)
  [arXiv:0809.3505 [hep-th]].
   

 \bibitem{tetradis} 
  N.~Tetradis,
  JHEP {\bf 1003}, 040 (2010)
  [arXiv:0905.2763 [hep-th]]. 
  
    
\bibitem{chamblin} 
  A.~Chamblin, R.~Emparan, C.~V.~Johnson and R.~C.~Myers,
  Phys.\ Rev.\ D {\bf 60}, 064018 (1999)
  [hep-th/9902170].
  
\bibitem{kajantie} 
  K.~Kajantie, J.~Louko and T.~Tahkokallio,
  Phys.\ Rev.\ D {\bf 78}, 126011 (2008)
  [arXiv:0809.4875 [hep-th]].

\bibitem{sin} 
  S.~-J.~Sin, S.~Nakamura and S.~P.~Kim,
  JHEP {\bf 0612}, 075 (2006)
  [hep-th/0610113].
  
  \bibitem{fischetti} 
  S.~Fischetti, D.~Kastor and J.~Traschen,
  arXiv:1407.4299 [hep-th].
  
  
  
  \bibitem{horowitz} 
  G.~T.~Horowitz and S.~F.~Ross,
  JHEP {\bf 9804}, 015 (1998)
  [hep-th/9803085].
  
  \bibitem{witten2} 
  E.~Witten,
  Adv.\ Theor.\ Math.\ Phys.\  {\bf 2}, 505 (1998)
  [hep-th/9803131].
  
  
  \bibitem{erlich2} 
  J.~Erlich, E.~Katz, D.~T.~Son and M.~A.~Stephanov,
  Phys.\ Rev.\ Lett.\  {\bf 95}, 261602 (2005)
  [hep-ph/0501128].
  
   \bibitem{pisarski2} 
  R.~D.~Pisarski and M.~Tytgat,
  In *Minneapolis 1996, Continuous advances in QCD* 196-205
  [hep-ph/9606459]. 
 
   \bibitem{beringer}
J.~Beringer et al. (Particle Data Group), Phys.\ Rev.\ D {\bf 86}, 010001 (2012).


%


%
%
 
%
%

    
%

 

\end{thebibliography}
\end{document}